# Possibility of a (bi)polaron high-temperature superconductivity in Poly A / Poly T DNA duplexes.

V. D. Lakhno, V.B. Sultanov

Institute of Mathematical Problems of Biology RAS,
142290, Pushchino, Moscow region, Russia.

Abstract.

Dynamical equations for a polaron and bipolaron in a DNA duplex are obtained on the basis of Holstein-Hubbard Hamiltonian. It is shown that in a PolyA/PolyT duplex especially stable is a bipolaron state in which holes are localized on different chains of the duplex. With the use of the polaron and bipolaron free energy, the temperature of bipolaron decay is found to be $T_d \approx 350$ K which can serve as an approximate estimate of the superconducting transition temperature. The way of constructing superconducting nanowires on the basis of DNA is suggested.

Key words: DNA duplex, electron, hole, bipolaron, high-temperature superconductivity.

## I. Introduction.

Ideas that organic molecules can possess superconductivity were suggested as early as in the middle of the last century [1, 2]. As applied to DNA, these ideas based on BCS theory were used in [3, 4]. Notwithstanding the fact that superconductivity in organic molecules was discovered long ago, superconducting properties of DNA have been found quite recently [5, 6].

Up to now, the question of the mechanism of charge transfer in DNA is still unclarified. Experiments demonstrate that DNA can be a dielectric [7, 8], a metal [9, 10], a semiconductor [11-13] and, as was shown above, a superconductor [5, 6]. The role of polarons in the charge transfer process, which represent particle-like states arising as a result of an interaction between charge carriers (holes, as a rule) and oscillations of nucleotide pairs is pointed out in [14-18]. The fact that DNA can demonstrate superconducting properties shows that charge carriers may be holes in a bipolaron state. According to BCS theory [19], in ordinary superconductors, pairs of electrons with opposite momentums and spins, when interacting with phonons, form bound states whose Bose condensate in the momentum space provides superconductivity. In explaining the DNA superconductivity we based our research on the model in which bound states, or bipolarons, form pairs of polarons with opposite spins of particles which leads to a Bose condensate with pairing in real space [20-23].

Earlier [24] we constructed such a model for the case of a PolyG/PolyC duplex in a single-chain approximation. In [25] the theory was generalized to the case when the holes were able to move along both the chains of the duplex in the approximation of independent oscillators: each nucleotide corresponded to its own oscillator.

In this paper we use a more realistic model (section 2) in which an oscillator is put into correspondence not to an individual nucleotide, but to a Watson-Crick pair.

In DNA, a superconducting state was observed at temperature $T \approx 1$ K for aperiodic nucleotide sequences ($\lambda$-DNA) [5, 6]. In this case Bose condensation of a bipolaron gas is hardly realizable because of inhomogeneity of a nucleotide sequence. This may be a reason why the temperature of superconducting transition is low. Bipolarons in homogeneous nucleotide sequences can be more stable than in aperiodic ones. Here we will consider the conditions under which bipolaron states can form in homogeneous Poly G / Poly C and Poly A / Poly T DNA synthetic duplexes.

## II. Physical model.

Recall that DNA consists of four types of nucleotides denoted by the type of their constituent nitrogenous bases as A (adenine), T (thymin), C (cytosine) and G (guanine) which unite into complementary (Watson-Crick) pairs so that nucleotide A always pairs with T and nucleotide G always pairs with C. Two polymer chains pairwise coupled by hydrogen nucleotide bonds form a double helix. Now rather long duplexes with preset sequences of nucleotide pairs can be synthesized artificially. Of great interest are homogeneous duplexes which can serve as molecular wires in nanoelectronic devices [26]. In most experiments on charge transfer in DNA a charge is transferred not by electrons but by holes [26, 27]. If a nucleotide loses an electron, the arising hole has a potential energy V such that $V_G < V_A < V_C < V_T$. In the simplified DNA model under consideration we believe that the planes of nucleotide bases are parallel to each other at each instant of time and the distances between the planes of bases of neighbouring sites are unvaried, i.e. we deal with a standard DNA model. Hence we think that when holes are inserted into a DNA double helix, the chains are deformed in the region of the hole occurrence along the direction of hydrogen bonds (Fig. 1).

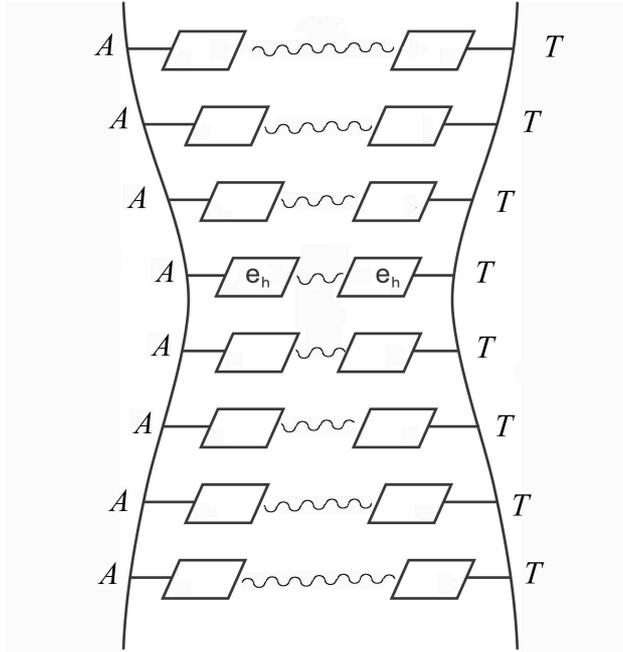

Fig. 1. Displacement of nucleotides along hydrogen bonds in the region of holes $e_h$ occurence in the bipolaron state.

Transfer of a hole in DNA is determined by overlapping of its wave functions on neighboring sites. If the transfer occurs along DNA chains, then, in view of the assumed symmetry, overlapping integrals along each chain are practically independent of the intrasite displacements of nucleotides irrespective of the displacements length. In this case overlapping integrals between the chains will depend on the sites displacements. However these integrals are less by an order of magnitude than those for neighbouring nucleotides on one chain. Hence, for small values of sites displacements a change in overlapping integrals will be small. It can be said that in our model maximum displacements of nucleotides occur along the direction of hydrogen bonds of each Watson-Crick pair. An additional argument in favour of the model is the fact that it is the limit case of the Peyrard-Bishop-Dauxois model which adequately describes dynamical properties of a DNA molecule [28].

**III. Mathematical model.**

Our consideration is based on a model of a double-stranded DNA which contains holes interacting with oscillations of nucleotide pairs (sites) considered as harmonic oscillators. Hamiltonian of the model, satisfying requirements of section 2, is a Holstein-Hubbard Hamiltonian having the form [29-31]:

$$\hat{H} = \sum_{(i,j),\sigma} \eta_{ij} c^+_{i\sigma} c_{j\sigma} + \sum_n \hbar\omega \left( a^+_n a_n + \frac{1}{2} \right) + \sum_n g\hat{n}_n \left( a^+_n + a_n \right) + \sum_j U\hat{n}_{j\uparrow}\hat{n}_{j\downarrow}, \qquad (1)$$

where $\hat{n}_{j\sigma} = c^+_{j\sigma} c_{j\sigma}$, $\hat{n}_n = \hat{n}_{2n-1\uparrow} + \hat{n}_{2n-1\downarrow} + \hat{n}_{2n\uparrow} + \hat{n}_{2n\downarrow}$, j = 1,...,2N is the nucleotide number, n = 1,...,N is the site number (see Fig. 2), $a^+_n$, $a_n$ are operators of the birth and annihilation of a phonon on the n-th site, $c^+_{j\sigma}$, $c_{j\sigma}$ are operators of the birth and annihilation of a hole with spin σ on the j-th nucleotide, $\eta_{ij}$ is a matrix element of the hole transition between neighboring nucleotides (i, j), g is the constant of interaction between a hole and oscillations of nucleotide pairs, U is the Coulomb repulsion parameter, ℏω is the energy of stretch oscillations of nucleotide pairs (oscillations along hydrogen bonds in a nucleotide pair) for which the interaction constant g is much greater than for other types of oscillations [32]. In the case of DNA, our model is physically equivalent to a Holstein molecular chain where each site contains a nucleotide pair rather than an atom pair. As distinct from the model in [25], here displacements of an oscillator on the site are affected by the total density of the inserted charges distributed on two nucleotides which leads to a very stable bipolaron on the A/T duplex; see equations (7) and (9) for polaron and bipolaron wave functions below.

For an arbitrary DNA duplex containing N nucleotide pairs, we will use the numeration of nucleotides and sites (let the upper chain have the direction 5'→3') shown in Fig 2:

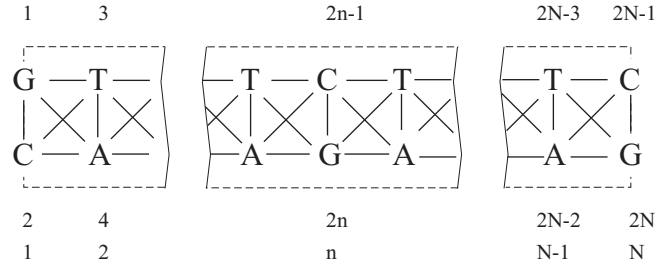

Fig. 2. Numeration of nucleotides (j=1÷2N); at the bottom is the numeration of sites (n=1÷N).

Then the electronic part of the Hamiltonian (1), which takes account of all the hole transitions between neighboring nucleotides, including those between nucleotides on complementary chains, can be written in a block tridiagonal form with blocks – matrices of the second order:

$$\hat{H}_e = \begin{pmatrix} D_1 & B_1 & 0 & ... & 0 & 0 \\ B_1^* & D_2 & B_2 & ... & 0 & 0 \\ 0 & B_2^* & D_3 & ... & 0 & 0 \\ ... & ... & ... & ... & ... & ... \\ 0 & 0 & 0 & ... & D_{N-1} & B_{N-1} \\ 0 & 0 & 0 & ... & B_{N-1}^* & D_N \end{pmatrix}, \quad D_n^* = D_n.$$

Diagonal elements of $\hat{H}_e$ are the potentials of nucleotide oxidation [33], non-diagonal elements are exchange integrals between one-electron orbitals of neighboring nucleotides [34, 35]. In the following we use matrix elements of Hamiltonian $\hat{H}_e$ expressed in eV.

In the case of homogeneous duplexes $D_n = \begin{pmatrix} 0 & d_{12} \\ d_{12} & d \end{pmatrix}$, $B_n = \begin{pmatrix} b_1 & b_{12} \\ b_{21} & b_2 \end{pmatrix}$ (0 stands for the lowest oxidation potential of two nucleotides). According to [33-35], for the duplex G/C $D_n = \begin{pmatrix} 0 & 0.05 \\ 0.05 & 0.66 \end{pmatrix}$, $B_n = \begin{pmatrix} 0.084 & 0.01 \\ 0.025 & 0.041 \end{pmatrix}$; for the duplex A/T $D_n = \begin{pmatrix} 0 & 0.034 \\ 0.034 & 0.21 \end{pmatrix}$, $B_n = \begin{pmatrix} 0.03 & 0.016 \\ 0.007 & 0.158 \end{pmatrix}$ (Fig. 3).

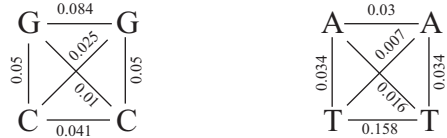

Fig. 3. Matrix elements of the hole transitions in G/C and A/T duplexes.

Notice that as N→∞ the spectrum of allowed energies of the electronic Hamiltonian $\hat{H}_e$ represents two closed segments. For the duplex G/C these are segments ≈ [-0.166, 0.154] and ≈ [0.58, 0.75], which are strongly separated. For the duplex A/T the segments intersect. Fig. 4 shows the dependence of the points of the $\hat{H}_e$ spectrum on the matrix element $b_2$ (N=11). In the case of a real value of $b_2$=0.158 the band of T completely absorbs the band of A ($V_A$=0.45 < $V_T$=0.66).

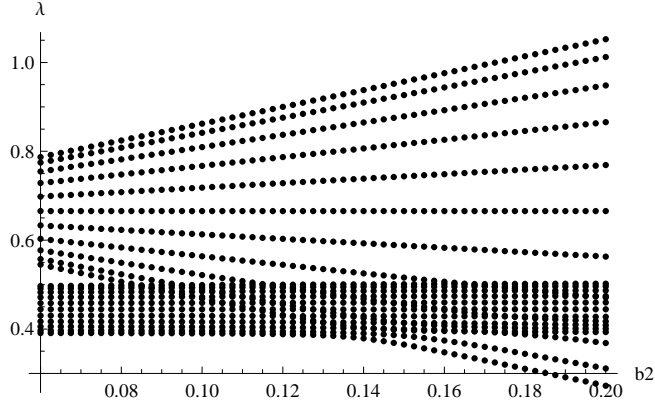

Fig. 4. Dependence of the spectrum of the electronic Hamiltonian $\hat{H}_e$ on $b_2$ for A/T duplex.

This difference in the spectra leads to considerably different properties of both polarons and bipolarons on G/C and A/T duplexes (see below).

**IV. Polarons at temperature T = 0.**

If the system contains only one hole and U=0, then spin indices in Hamiltonian (1) can be omitted. Let us pass on in Hamiltonian (1) from operators of birth and annihilation of phonons $a_n^+$, $a_n$ to coordinates $Q_n$:

$$a_n^+ = \frac{1}{\sqrt{2}}\left(-\frac{d}{dQ_n}+Q_n\right), \quad a_n = \frac{1}{\sqrt{2}}\left(\frac{d}{dQ_n}+Q_n\right). \quad (2)$$

As a result, for a polaron in a duplex, we will have a Hamiltonian:

$$\hat{H} = \sum_{i,j}\eta_{ij}c_i^+c_j + \alpha\sum_n\left(c_{2n-1}^+c_{2n-1}+c_{2n}^+c_{2n}\right)q_n + \sum_n\left(\frac{\hat{p}_n^2}{2M}+\frac{K}{2}q_n^2\right), \quad (3)$$

where $q_n = \sqrt{\frac{\hbar}{M\omega}}Q_n$, $\alpha = g\sqrt{\frac{2M\omega}{\hbar}}$, $\hat{p}_n = -i\hbar\frac{\partial}{\partial q_n}$, M is the effective mass of a nucleotide pair, K=Mω² is the elastic constant. Let us present the wave function $|\Psi\rangle$ of a polaron corresponding to Hamiltonian (1) as an expansion in coherent states:

$$|\Psi\rangle = \sum_n\left(\psi_{2n-1}c_{2n-1}^+ + \psi_{2n}c_{2n}^+\right)a_n^+\exp\left\{-\frac{i}{\hbar}\sum_m[\beta_m(t)\hat{p}_m - \pi_m(t)q_m]\right\}|0\rangle, \quad (4)$$

where $|0\rangle$ is the vacuum wave function, $\beta_n(t) = \langle\Psi(t)|q_n|\Psi(t)\rangle$, $\pi_n(t) = \langle\Psi(t)|\hat{p}_n|\Psi(t)\rangle$.
With the use of (4), the total energy of the system $E = \langle\Psi|\hat{H}|\Psi\rangle$ will be written as:

$$E = \sum_{i,j}\eta_{ij}\psi_i^*\psi_j + \alpha\sum_n \left(|\psi_{2n-1}|^2 + |\psi_{2n}|^2\right)\beta_n + \sum_n\left(\frac{\pi_n^2}{2M} + \frac{K}{2}\beta_n^2\right). \tag{5}$$

Hamiltonian equations take on the form

$$i\frac{d\psi_j}{dt} = \left(\hat{H}_e\vec{\psi}\right)_j + \alpha\beta_n\psi_j, \tag{$6_1$}$$

$$\frac{d^2\beta_n}{dt^2} = -K\beta_n - \alpha\left(|\psi_{2n-1}|^2 + |\psi_{2n}|^2\right), \tag{$6_2$}$$

where j=2n-1, 2n, n=1,…,N. In the steady state from ($6_1$, $6_2$) follows a stationary Discrete Nonlinear Schroedinger Equation (DNSE) for the polaron wave function $\vec{\psi} = \{\psi_j\}_{j=1}^{2N}$

$$\left(\hat{H}_e\vec{\psi}\right)_j - \frac{\kappa}{2}\left(|\psi_{2n-1}|^2 + |\psi_{2n}|^2\right)\psi_j = \lambda\psi_j, \text{ j=2n-1,2n; n=1,…,N.}$$

(7)

Notice that in the steady case the form of equations ($6_1$, $6_2$) is close to that obtained for high – $T_c$ superconductors [36].

The value of $\kappa = \frac{4g^2}{\hbar\omega}$ determines the interaction of a charge with sites oscillations and for DNA is equal to κ=0.5267 eV [37]. The ground state is determined by minimization of the total energy functional $\Phi_p(\vec{\psi}) = \sum_j\left(\hat{H}_e\vec{\psi}\right)_j\psi_j - \frac{\kappa}{4}\sum_n\left(|\psi_{2n-1}|^2 + |\psi_{2n}|^2\right)^2$, the normalizing condition being $\sum_j|\psi_j|^2 = 1$. Fig. 5 presents the dependence of the polaron energy $E_p = Min[\Phi_p(\vec{\psi})]$ on the G/C duplex on κ/2 :

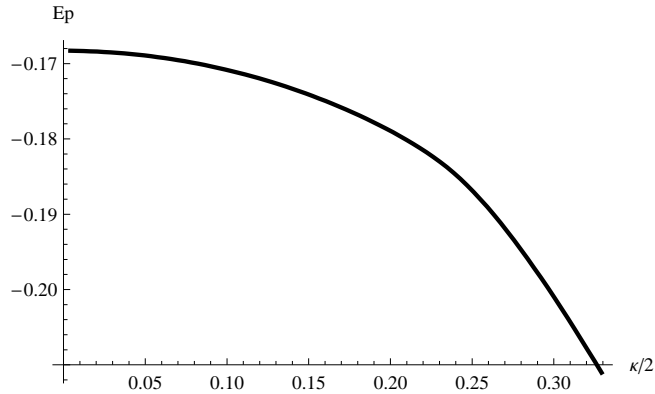

Fig. 5. Polaron energy on G/C duplex.

On the G/C duplex one observes a continuous transition from a big polaron to a small one as κ increases, the polaron is concentrated on the chain G, the population density of the chain C is negligible (Fig. 6).

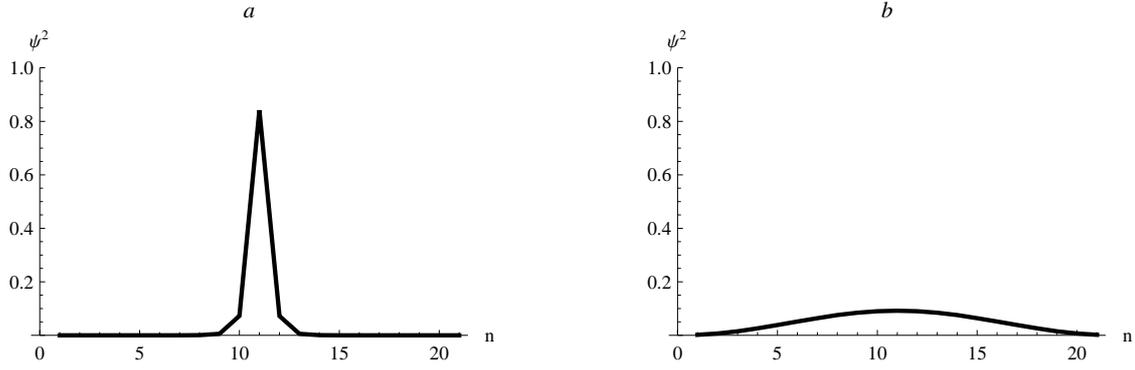

Fig. 6. Polarons on the duplex G/C, N=21, a: κ/2=0.33, b: κ/2=0.004.

Fig. 7 shows the dependence of the polaron energy $E_p = Min[\Phi_p(\vec{\psi})]$ on the A/T duplex on κ/2:

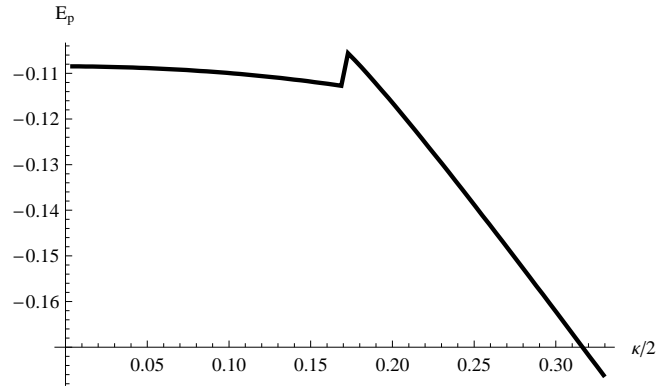

Fig. 7. Polaron energy on A/T duplex.

On the A/T duplex, there is a leap for a certain value of κ. At the leap point the ground state changes qualitatively which is called autolocalization. This phenomenon is associated with the structure of the Hamiltonian spectrum. For large κ, the polaron is localized and nearly concentrated on the chain A (in Fig. 8a, κ/2=0.173, the population densities are: $(A)_N$ =0.76, $(T)_N$ =0.24), for some κ, the polaron jumps onto the chain T and is delocalized (in Fig. 8b, κ/2=0.169 the population densities are: $(A)_N$ =0.06, $(T)_N$ =0.94).

It can be shown numerically that at the point of jump, as κ decreases, the localized state loses its stability, or no longer remains the total energy minimum.

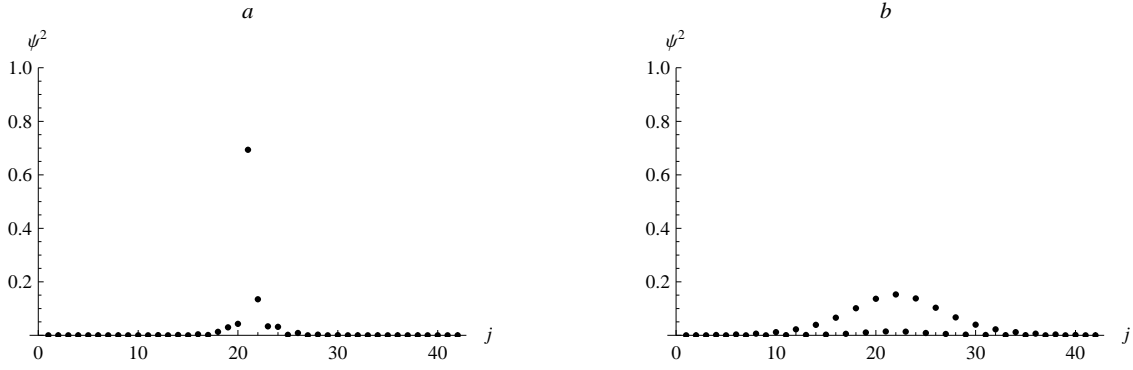

Fig. 8. Polarons on the duplex A/T, N=21, a: κ/2=0.173, b: κ/2=0.169.

Interestingly that for the A/T duplex (Fig. 7), the dependence of the polaron energy on the coupling constant qualitatively coincides with the same dependence in 2D $CuO_2$ planes of high-temperature superconductors obtained in [36]. This fact testifies that, as distinct from G/C duplexes which are actually 1D electron systems, A/T duplexes are identical to 2D systems in their electron properties.

## V. Bipolarons at temperature T = 0.

In a one-dimensional case, irrespective of a particular type of a molecular chain, bipolaron states for Hamiltonian (1) were considered in paper [31] in the anticontinious limit, i.e. for small values of the matrix element η.

In the case of a bipolaron the wave function of the ground state of Hamiltonian (1) for DNA duplex has the form:

$$|\Psi\rangle = \sum_{n,m} \left( \psi_{2n-1,2m-1} c^+_{2n-1\uparrow} c^+_{2m-1\downarrow} + \psi_{2n,2m-1} c^+_{2n\uparrow} c^+_{2m-1\downarrow} + \psi_{2n-1,2m} c^+_{2n-1\uparrow} c^+_{2m\downarrow} + \psi_{2n,2m} c^+_{2n\uparrow} c^+_{2m\downarrow} \right) \times$$

$$\times a^+_n \exp\left\{ -\frac{i}{\hbar} \sum_n \beta_n(t)\hat{p}_n - \pi_n(t) q_n \right\} |0\rangle ,$$

where $\hat{p}_n$ and $q_n$ are determined by (3). Accordingly, the total energy $E = \langle \Psi | \hat{H} | \Psi \rangle$ of the system is equal to

$$E = \sum_{i',j,i} \eta_{i'i} \psi^*_{i'j} \psi_{ij} + \sum_{i',j,i} \eta_{i'j} \psi^*_{ii'} \psi_{ij} + \alpha \sum_n \sum_j \left( |\psi_{2n-1,j}|^2 + |\psi_{2n,j}|^2 + |\psi_{j,2n-1}|^2 + |\psi_{j,2n}|^2 \right) \beta_n + U \sum_j |\psi_{jj}|^2 +$$

$$+ \sum_n \left( \frac{\pi^2_n}{2M} + \frac{K}{2} \beta^2_n \right).$$

This expression yields the following dynamical equations for the bipolaron:

$$i\frac{d\psi_{ij}}{dt} = \left(\hat{H}_e\vec{\psi}_{(i)} + \hat{H}_e\vec{\psi}_{(j)}\right)_{ij} + \left(\alpha(\beta_n + \beta_m) + U\delta_{ij}\right)\psi_{ij}, \quad i=2n-1, 2n;\ j=2m-1, 2m,\ n,m=1,..,N,$$

(8$_1$)

$$M\frac{d^2\beta_n}{dt^2} = -K\beta_n - \alpha\sum_j\left(|\psi_{2n-1,j}|^2 + |\psi_{j,2n-1}|^2 + |\psi_{2n,j}|^2 + |\psi_{j,2n}|^2\right), \tag{8$_2$}$$

Hamiltonian $\hat{H}_e$ acts on the 1-st and the 2-d coordinates of the wave function $\vec{\psi} = \{\psi_{ij}\}_{i,j=1}^{2N}$.

In the steady state from (8$_1$, 8$_2$) we get a steady DNSE for the bipolaron wave function determining the state of a pair of holes with opposite spins:

$$\left(\hat{H}_e\vec{\psi}_{(i)} + \hat{H}_e\vec{\psi}_{(j)}\right)_{ij} + \left(-\kappa(\rho_n + \rho_m) + U\delta_{ij}\right)\psi_{ij} = \lambda\psi_{ij}, \quad i=2n-1, 2n;\ j=2m-1, 2m;\ n, m=1,\ldots,N,$$

(9)

which is similar to equation (7) for the polaron wave function $\rho_n = \frac{1}{2}\sum_k\left(|\psi_{2n-1,k}|^2 + |\psi_{k,2n-1}|^2 + |\psi_{2n,k}|^2 + |\psi_{k,2n}|^2\right)$. It is easy to verify that equation (9) changes into equation for the bipolaron wave function on a single-strand chain if we put $\psi_{ij} = 0$ for even indices i or j.

The ground state is determined by minimization of the functional

$$\Phi_{bp}(\vec{\psi}) = \sum_{i,j}\left(\hat{H}_e\vec{\psi}_{(i)} + \hat{H}_e\vec{\psi}_{(j)}\right)_{ij}\psi_{ij} + U\sum_i|\psi_{ii}|^2 - \frac{\kappa}{2}\sum_{i,j}(\rho_n + \rho_m)|\psi_{ij}|^2$$

under condition of symmetry $\psi_{ij} = \psi_{ji}$ and normalization $\sum_{i,j}|\psi_{ij}|^2 = 1$. The ground state energy is equal to $E_{bp} = Min[\Phi_{bp}(\vec{\psi})]$.

In the case under study, when the charged particles are holes, the parameter $U \neq 0$. Its exact value is unknown but approximate value can be obtained from the following considerations. For two holes localized on one nucleotide, the energy of Coulomb repulsion is roughly equal to $e^2/2\varepsilon_\infty\ell$, where $\varepsilon_\infty$ is a high-frequency dielectric permittivity of a molecule, $\ell$ is a characteristic size of a nucleotide. For DNA, these quantities have the order of magnitude $\varepsilon_\infty = 2$ [38], $\ell = 3$Å, , which yields for U the value U≈1 eV. For this value of U, localization of two holes on one nucleotide is impossible. A lower energy value is achieved when holes are localized on two or more nucleotides.

The estimates U≈1eV, κ≈0.5 eV are approximate. Moreover, the value of κ can vary widely depending on external conditions. The value κ≈0.5 eV corresponds to a "dry"

DNA molecule. In a solution the value of κ can considerably enhance (according to [39, 40] approximately 5 times). In this case the conditions of bipolaron formation will more likely be created.

The condition of stability of bipolaron states for T=0 is determined by the inequality:

$$E_{bp} < 2E_p. \qquad (10)$$

Bipolaron and polaron states were calculated by minimization of the functionals $\Phi_{bp}(\psi_{ij})$ and $\Phi_p(\psi_i)$ when symmetry and normalization conditions were fulfilled for the parameters κ, U, in the region containing their values for DNA mentioned above. Fig. 9 shows the diagram of bipolaron states stability determined by inequality (10) on G/C duplex.

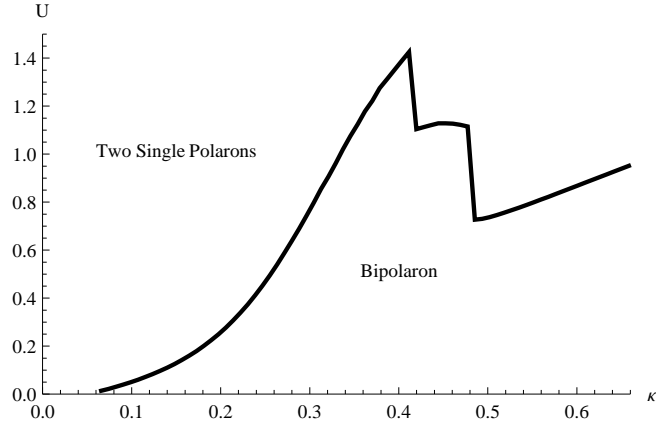

Fig. 9. Region of bipolaron stability on G/C duplex.

As is seen from Fig. 9, the values of parameters κ=0.5267 eV and U=0.75 eV lie in the region of bipolaron stability on G/C duplex. The energy of bipolaron binding at these parameters is equal to $\Delta = E_{bp} - 2E_p \approx -1.5 \cdot 10^{-4}$ eV.

Fig. 10 demonstrates charge densities $p_n = 2\sum_i |\psi_{i,2n-1}|^2$ of bipolarons on the boundary of the stability region $E_{bp} = 2E_p$ on the chain $(G)_N$, N=21, population density of the chain $(C)_N$ is less than 0.003.

Fig. 10 suggests that in the stability region, the bipolaron bound state is formed by polarons, localized on sites, separated by 0, 1, 2, …, 6 intersite distances. Notice that in the case of U=0 the bipolaron state is stable for any parameter values, the state with the lowest energy corresponding to a two-particle state localized on one nucleotide. The bipolaron state with U=0 could correspond to Frenkel biexiton self-localized states. As U=∞, i.e. when the holes are separated by a large distance, the bipolaron state on G/C duplex changes into two separate polaron states.

It is well known that in adiabatic limit of the Holstein model holes (electrons) collapse into autolocalized polarons or bipolarons due to spontaneous breaking of translational invariance at $\frac{1}{4}(\hbar\omega/2g)^4 \ll 1$. Since autolocalized states can form with equal probability at any site of the chain they appear to be energetically degenerated and form polaron or bipolaron energy bands $\Delta E_p$, $\Delta E_{bp}$, restoring thereby a translational symmetry lost during the formation of a localized state.

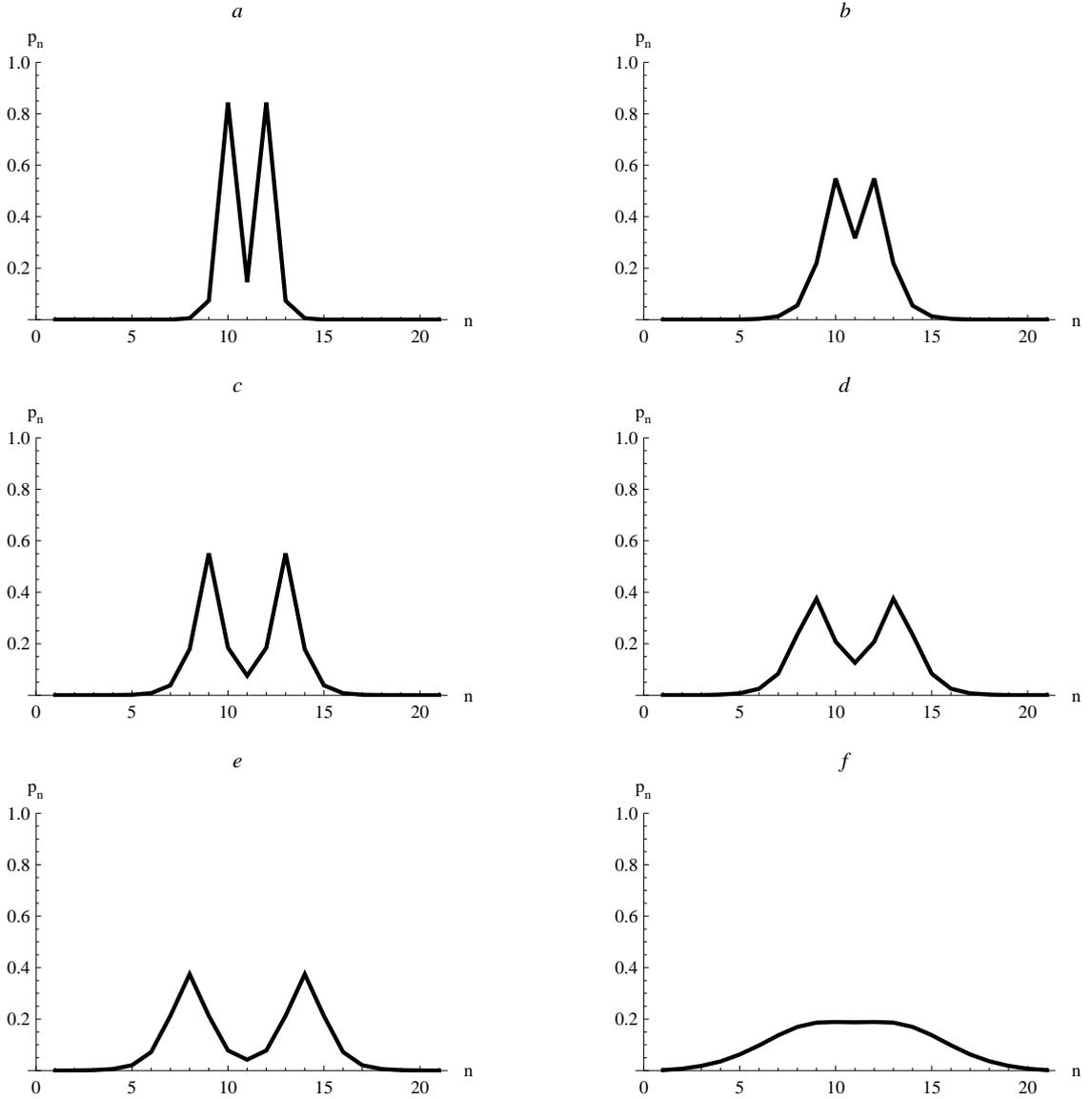

Fig. 10. Charge densities of bipolarons on the G/C duplex, a: κ=0.66, U=0.95; b: κ=0.49, U=0.73; c: κ=0.48, U=1.115; d: κ=0.42, U=1.104; e: κ=0.41, U=1.425; f: κ=0.165, U=0.16.

On the A/T duplex bipolarons appear to be very stable: inequality (10) is fulfilled everywhere over the considered region of κ, U parameters. For κ=0.5267 eV and U=1 eV the bipolaron binding energy is equal to $\Delta = E_{bp} - 2E_p \approx$ -0.13 eV. The relevant bipolaron is localized on the site in the center of the duplex, on nucleotides $A_N$, $T_{N+1}$ (Fig. 11). A bipolaron with so great coupling energy and small mass (apex perovskite) was obtained in paper by A. S. Alexandrov [41] where the estimation of the superconducting transition temperature $T_c \approx$ 300 K was made.

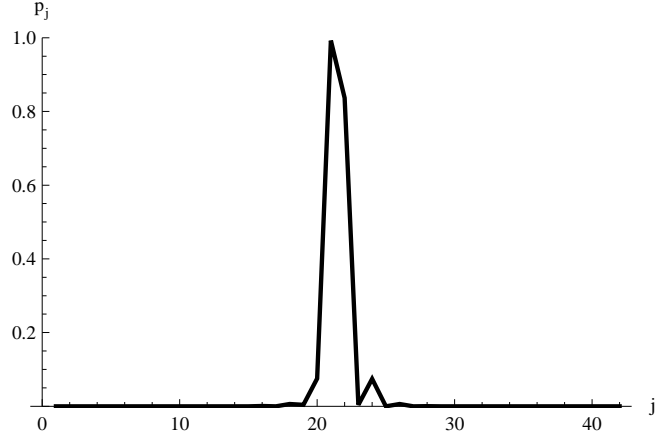

Fig. 11. Charge density $p_j$ of the bipolaron on the A/T duplex (N=21).

## VI. Bipolarons at temperature T ≠ 0.

Irrespective of a particular mechanism of superconductivity that arises in DNA due to polaron and bipolaron states, let us assess approximately the bipolaron decay temperature $T_d$ based on the bipolaron stability condition. At temperature T > 0 let the bipolaron stability conndition be:

$$F_{bp}(T) < 2F_p(T), \qquad (11)$$

where $F(T) = -T \ln Z(T)$ is the free energy of the bipolaron and polaron, respectively, calculated from the semiclassical statistical sum

$$Z(T) = \int ... \int \prod_n d\beta_n \left( \sum_j e^{-\lambda_j/T} \right) e^{-\sum_n \frac{K}{2}\beta_n^2/T},$$

where $K = 0.062$ eV/Å$^2$ [37], $\lambda_j = \lambda_j(\beta_1,...,\beta_N)$ are eigen values of Hamiltonians of the bipolaron and polaron, respectively for the preset displacements of oscillators $\{\beta_n\}$.

The integral is calculated by the Monte-Carlo method as a mean of the subintegral function from a random sample of points $\{\beta_n\}$ uniformly distributed on hypercube $[-R, R]^N$. For each point $\{\beta_n\}$ one solves a linear eigenvalue problem which is a 2N-dimensional for the polaron and $(2N)^2$ –dimensional for the bipolaron. The eigenvalue equations for the polaron are:

$$\left(\hat{H}_e \vec{\psi}\right)_j + \alpha \beta_n \psi_j = \lambda \psi_j, \quad j=2n-1, 2n; \; n=1,\ldots,N,$$

accordingly, those for the bipolaron are:

$$\left(\hat{H}_e \vec{\psi}_i + \hat{H}_e \vec{\psi}_j\right)_{ij} + \left(\alpha(\beta_n + \beta_m) + U\delta_{ij}\right)\psi_{ij} = \lambda \psi_{ij}, \quad i=2n-1, 2n; \; j=2m-1, 2m; \; n, m=1,\ldots,N,$$

where α=0.13 eV/Å [37], U=0.75 eV for G/C duplex and U=1 eV for A/T duplex.

Fig. 12 presents the calculation of $F_{bp}(T)$ and $2F_p(T)$ for G/C and A/T duplexes, N=5, R=5, the sample $\approx 10^4$ points.

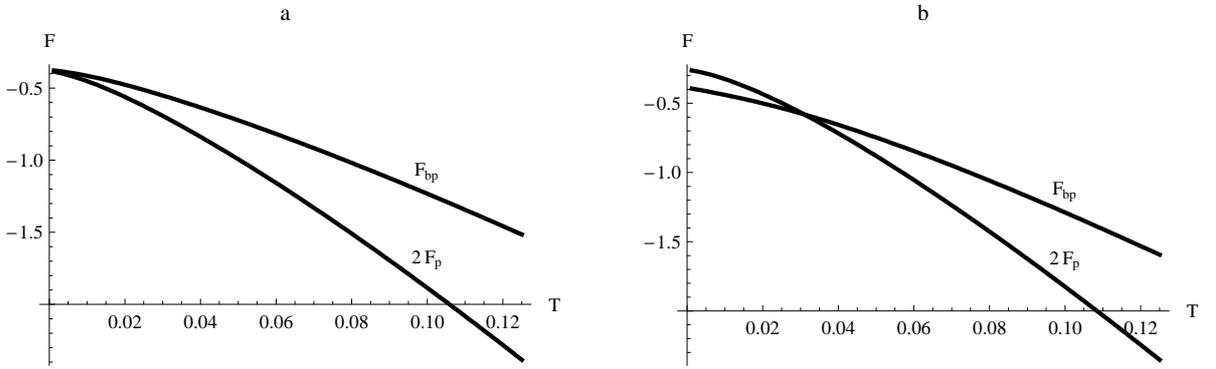

Fig. 12. Bipolaron binding energy a: G/C duplex, b: A/T duplex.

Fig. 12 suggests that for the G/C duplex, inequality (11) is not fulfilled as $T_d \approx 0$, for the A/T duplex, inequality (11) fails as $T_d \approx 0.03$ eV $\approx 350$ K. As $T = T_d$ the equality $F_{bp}(T_d) = 2F_p(T_d)$ is equivalent to $Z_{bp}(T_d) = Z_p^2(T_d)$, i. e. a bipolaron breaks down into two individual polarons.

The bipolaron decay temperature $T_d$ can serve as an approximate estimate of the superconducting transition temperature $T_c$. Indirectly it is testified by strong difference in decay temperatures for G/C and A/T duplexes.

As T→0 the value of the binding energy $\Delta(0) = F_{bp}(0) - 2F_p(0) = E_{bp} - 2E_p$ is consistent with that calculated above (see section 5) from the energy of the bipolaron and polaron ground state $\Delta = -1.5 \cdot 10^{-4}$ eV for the duplex G/C and $\Delta = -0.13$ eV for the duplex A/T.

**VII. Possibility of experimental verification of the theory.**

Notice, that earlier the possibility of bipolaron states in short (six-sites) heterogeneous oligonucleotide chains was pointed out in [42]. According to [42], in such chains pairs should localize on doublets, triplets and quadruplets of guanines.

Our model of bipolaron states in DNA is close to the theoretical model by Anderson [43] in which electrons in a pair are localized on neighbouring sites (for Poly A / Poly T duplex – on adenine A and thymine T of one of Watson-Crick pairs). In this case Coulomb repulsion appears to be rather weak which leads to formation of a very stable bipolaron which decays at temperature of $T_d \cong 350$ K which approximately estimates the superconducting transition temperature. It should be stressed that the estimate of $T_c$ obtained here is an upper bound. A real value of $T_c$ can be much smaller. For small-radius bipolarons $T_c$ will be determined by the width of the bipolaron band (i.e. bipolaron effective mass). In the case of a narrow band it will be very small. (see, however [44-46]). In the opposite limit of large-radius polarons the condition of a bipolaron decay into individual polaron states yields the upper estimate given above. In any case the estimates obtained testify that up to physiological temperatures electronic and optical properties of DNA will be determined by bipolaron states.

The main problem concerned with the possibility of a superconducting state in DNA is that charge carriers (electrons or holes) must occur there, since DNA itself is a dielectric. In experiment [5], when measuring conductivity, a DNA molecule was fastened to two rhenium-carbon electrodes residing on a micaceous substrate. The gap between the electrodes in [5] was carried out by burning-out a ≈0,5 μc thick band which represented a mica with islets of residual rhenium-carbon atoms not bonded to one another. In our opinion, the appearance of charge carriers in DNA could be caused by a contact of the molecule fastened to the substrate with rhenium-carbon islets of the gap.

Since at low temperatures rhenium is a superconductor, the authors of [5] believed that the superconductivity observed was induced. To prove that in [5] DNA superconductivity did take place the authors of [6] remade that experiment with platinum electrodes which do not turn into a superconducting state at any temperature and reproduced the results of [5]. This opens up possibilities for construction of superconducting PolyA / PolyT nanowires through electrolytic deposition of atoms onto the surface of PolyA/PolyT duplexes so that the atoms would serve as an acceptor admixture for p-duplexes and a donor admixture for n-duplexes, by analogy with p and n types of semiconductors.

The work was done with the support from the RFBR, projects № 11-07-12054-ofi-M-2011; 10-07-00112.

References

[1] F. London, Superfluids(John Wiley&Sons, Inc., New York, 1950), 1.

[2] W.A. Little, Phys. Rev. 134, (1964), A1416-A1424.

[3] J. Ladik, G. Biczo, J. Redly, Phys. Rev. 188 (1969) 710.

[4] J. Ladik, A. Bierman, Physics Lett., 29A (1969) 636.

[5] A.Y. Kasumov et al, Science 291 (2001) 280.

[6] A.D. Chepelianskii et al, New Journal of Physics, 13 (2011) 063046.

[7] A.J. Storm, J. Van Noort, S. De Vries, C. Dekker, Appl. Phys. Lett. 79 (2001) 3881.

[8] P.J. De Pablo et al., Phys. Rev. Lett. 85 (2000) 4992.

[9] F.-W. Fink, C. Schönenberg, Nature 398 (1999) 407.

[10] Y. Okahata, T. Kobayashi, K. Tanaka, M. Shimomura, J. Am. Chem. Soc. 120(1998) 6165.

[11] D. Porath, A. Bezryadin, S. De Vries, C. Dekker, Nature 403 (2000) 635.

[12] L. Cai, H. Tabata, T. Kawai, Appl. Phys. Lett. 77 (2000) 3105.

[13] K.-H. Yoo et al., Phys. Rev. Lett. 87 (2001) 198102.

[14] E.M. Conwell, S.V. Rakhmanova, Proc. Nat. Acad. Sci. USA 97(2000) 4556.

[15] E.M. Conwell, Proc. Nat. Acad. Sci. USA 102 (2005) 8795.

[16] C.-S. Liu, G.B. Shuster, J. Am. Chem. Soc. 125 (2003) 6098.

[17] V.D. Lakhno, J. Biol. Phys. 26 (2000) 133.

[18] V.D. Lakhno, Phys. Particles & Nucl. Letters 5 (2008) 231.

[19] J.R. Schrieffer, Theory of Superconductivity. New York: W.A.Benjamin, 1964.

[20] A.S. Alexandrov, J. Ranninger, Phys. Rev. B, 24 (1981) 1164.

[21] A.S. Alexandrov, JETP Lett., 46 (S1) (1987) S107.

[22] A.S. Alexandrov, J. Ranninger, S. Robaszkiewicz, Phys. Rev. B, 33 (1986) 4526.

[23] N.I. Kashirina, V.D. Lakhno, Uspekhi Fizicheskikh Nauk, 180 (2010) 449.

[24] V.D. Lakhno, V.B. Sultanov, Biophysics, 56 (2011), 210-213.

[25] V.D. Lakhno, V.B. Sultanov, Chem. Phys. Lett., 503 (2011), 292.

[26] V.D. Lakhno, Int. J. Quant. Chem. 108 (2008) 1970.

[27] A. Offenhäusser, R. Rinaldi (eds), Nanobioelectronics – for Electronics, Biology and Medicine 2009, Springer.

[28] T. Dauxois, M. Peyrard, A.R. Bishop, Phys. Rev. E, 47, (1993), R 44.

[29] T. Holstein, Ann. Phys. 8 (1959) 325.

[30] J. Hubbard, Proc. Roy. Soc. 276 (1963) 238.

[31] L. Proville, S. Aubry, Physica D 113 (1998) 307.

[32] E.B. Starikov, Phil. Mag. 85 (2005) 3435.

[33] F.D. Lewis, Y. Wu, J. Photochem. Photobiol. 2 (2001) 1.

[34] A.A. Voityuk, N. Rösch, M. Bixon, J. Jortner, J. Phys. Chem. B 104 (2000) 9740.

[35] A.A. Voityuk, J. Jortner, M. Bixon, N. Rösch, J. Chem. Phys. 114 (2001) 5614.

[36] V.V. Kabanov, O. Yu. Mashtakov, Phys. Rev. B 47 (1993) 6060.

[37] V.D. Lakhno Int. J. Quant. Chem. 110 (2010) 127.

[38] N.K. Balabaev, V.D. Lakhno, SPIE 1403 (1991) 478.

[39] A.A. Voityuk, J. Chem. Phys. 122 (2005) 204904.

[40] A.A. Voityuk, J. Phys. Chem. 109 (2005) 10793.

[41] A.S. Alexandrov, Superlight small bipolarons: a route to room temperature superconductivity. In "High-Tc Superconductors and Related Transition Metal Oxides" eds. A. Bussmann-Holder and H. Keller (Springer, 2007) 1-15.

[42] V. Apalkov, T. Chakraborty, Phys. Rev. B 73 (2006) 113103.

[43] P.W. Anderson, Science, 316, (2007) 1705.

[44] A.S. Alexandrov, P.E. Kornilovitch, Phys. Rev. Lett, 82 (1999) 807.

[45] Y.P. Hague, P.E. Kornilovitch, J.H. Samson, A.S. Alexandrov, Phys. Rev. Lett, 98 (2007) 037002.

[46] J.T. Devreese, A.S. Alexandrov, Rep. Progr. Phys. 72 (2009) 066501.